\documentclass[12pt]{article}
\usepackage{amsmath, amssymb, graphicx, float}
\usepackage{geometry}
\usepackage{physics}
\usepackage{hyperref}
\usepackage{caption}
\usepackage{subcaption}
\usepackage{cite}
\title{\textbf{Scaling Behavior in the Avalanche Activity Distribution of the Abelian Sandpile Model}}
\author{Anubhav Ganguly \\ Department of Physics, IISER Mohali}
\date{28th September 2025}

\begin{document}

\maketitle

\begin{abstract}
We study the scaling properties of the avalanche activity distribution in the 2D Abelian Sandpile Model (ASM). For any recurrent configuration $C$, let $N(R,R^{\prime})$ denote the number of topplings at site $R$ if we add a grain at site $R^{\prime}$. Then we define
\[
A(R)\ =\ \sum_{R^{\prime}}N(R,R^{\prime}),
\]
which gives the total number of topplings at site $R$ in an avalanche initiated from the configuration $C$, summed over all possible addition sites $R^{\prime}$. Numerically analyzing the probability distribution $P(A=a,L)$ over various lattice sizes $L$, we find that it follows the scaling form
\[
P(A=a,L)=\frac{1}{L^{2}}F\left(\frac{A}{L^{2}}\right),
\]
with $F(u)\sim u^{-1/2}$ for small $u\lesssim0.1$, $F(u)\sim e^{-c_3u-c_4u^2 }$for $u\gtrsim0.1$ .
\end{abstract}

\section{Introduction}

The Abelian Sandpile Model (ASM), introduced by Bak, Tang, and Wiesenfeld~\cite{BTW,Bak1988}, is a paradigmatic example of self-organized criticality (SOC). In SOC systems, a slowly driven dissipative system naturally evolves into a critical state without fine-tuning of parameters, exhibiting scale invariance, power-law distributions, and long-range correlations~\cite{Dhar1990,Dhar1999}. The term ``Abelian Sandpile Model'' was introduced later, and the model has since been extensively studied in both physics and mathematics~\cite{BakBook,RedigReview,GolesJarai,Pruessner}.

Majumdar and Dhar~\cite{Majumdar1992} established the equivalence between the ASM and the $q \to 0$ limit of the $q$-state Potts model, showing that the two-dimensional ASM can be described by a conformal field theory with central charge $c=-2$. This mapping allows exact determination of certain critical exponents, including the dynamic exponent $z = 5/4$ associated with the growth of avalanche fronts under the burning algorithm~\cite{Dhar1995}.

Studies of the ASM traditionally focus on avalanche statistics. Let $N(R',R)$ denote the number of topplings at site $R'$ when a grain is added at site $R$, starting from a stable configuration. The total avalanche size is
\[
s(R) = \sum_{R'} N(R',R).
\]
Much effort has gone into characterizing the asymptotic form
\[
\text{Prob}(s) \sim s^{-\tau},
\]
with extensive numerical and analytical work devoted to estimating the avalanche exponent $\tau$. While the exact value of $\tau$ remains elusive, it is well established that the mean avalanche size scales as $\langle s \rangle \sim L^2$~\cite{ShiloBiham}.

Analytical studies~\cite{Manna1991,Priezzhev1998} and numerical simulations~\cite{BenHur1996,ShiloBiham} indicated that the stochastic Manna model shares some universal features with the deterministic ASM introduced by Bak, Tang, and Wiesenfeld. However, extended numerical analyses of critical exponents, multifractal analyses~\cite{Ostojic2003,Levine2016}, and studies of sandpiles as closed systems~\cite{Frette1996,Tetzlaff2010} show that deterministic and stochastic models belong to distinct universality classes. 

Following the framework of Engsig and Sneppen~\cite{Engsig2025}, we focus on the complementary observable
\[
A(R) = \sum_{R'} N(R,R'),
\]
which measures the total activity at site $R$ when grains are added randomly, keeping the background configuration $C$ fixed. The total activity satisfies
\[
\sum_R A(R) = \sum_R s(R),
\]
so that $\langle A \rangle = \langle s \rangle$. The distribution of $A$ has not been systematically analyzed previously in literature.

We investigate the distribution $\text{Prob}(A,L)$ of site activities in the two-dimensional ASM. Unlike avalanche sizes, $\text{Prob}(A,L)$ tends to zero for fixed $A$ as $L \to \infty$, and follows a scaling form
\[
P(A=a,L) = \frac{1}{L^2} F\!\left(\frac{a}{L^2}\right),
\]
with distinct behavior for small $a$. Our numerical results show \(F(u) \sim u^{-1/2}\) for small \(u\) and \(F(u) \sim e^{-au-bu^2}\) for \(u \gtrsim 0.1\), with \(a \sim 10\) and \(b \sim O(1)\). These findings complement previous studies of extended SOC systems~\cite{Frette1996}, network-based SOC models~\cite{Tetzlaff2010}, neural SOC activity~\cite{Chialvo2010}, and recent experimental results~\cite{Helmrich2020,Plenz2021}.

The remainder of the paper is organized as follows. In Sec.~II, we define the ASM and the activity observable. Sec.~III describes the numerical simulations and data analysis. Sec.~IV presents results on activity scaling. Sec.~V discusses implications and broader connections.

\section{Model and Definitions}

We analyze the classical two-dimensional Bak–Tang–Wiesenfeld (BTW) Abelian sandpile model on a square lattice of size \( L \times L \), as described in . Each site \( (x, y) \) has an integer-valued height \( \Delta_{x,y} \in \{0, 1, 2, 3\} \) in the relaxed state, and \( \Delta_{x,y} \geq 4 \) indicates an unstable (excited) site.

A grain is added to a randomly selected site \( (x, y) \), so that:

\[
\Delta_{x,y} \rightarrow \Delta_{x,y} + 1.
\]

If \( \Delta_{x,y} \geq 4 \), the site relaxes by distributing one grain to each of its four neighbors:

\[
\Delta_{x,y} \rightarrow \Delta_{x,y} - 4, \quad
\Delta_{x\pm1,y} \rightarrow \Delta_{x\pm1,y} + 1, \quad
\Delta_{x,y\pm1} \rightarrow \Delta_{x,y\pm1} + 1.
\]

This toppling may trigger further relaxations. The process continues until all \( \Delta_{x,y} < 4 \)
Open boundary conditions are used: when a boundary site topples, excess grains exit the system. Only once the lattice is stable again (i.e., all \( \Delta_{x,y} < 4 \)), a new grain is added.

Given a stable configuration \( C \) of the lattice, and a site position \(\Vec{R}\), we define the \textit{excitability function} as
\[
s(R) = \text{Size of the avalanche triggered by adding a grain at } R.
\]

If the addition at \(R\) causes no topplings, then \(s(R) = 0\).
Sites with \(s(R) > 0\) are called \emph{excitable}, while those that trigger large avalanches—for example, \(s(R) > 0.1L^2\)—are considered \textit{highly excitable}.

To compute \(s(R)\), a grain is added at \(R\), the avalanche is simulated, 
and the system is then reset to the original configuration \(C\). 
This procedure is repeated for all sites.

The \textit{activity function} quantifies the total number of topplings 
that a site \(R\) undergoes when avalanches are initiated at all possible 
sites \(R \in [1,L] \times [1,L]\) on the same configuration \(C\):

The \textit{activity function} quantifies the total number of topplings that a site \(R\) undergoes when avalanches are initiated at all possible sites \(R \in [1,L] \times [1,L]\) on the same configuration \(C\).

\[
A(R) = \sum_{R'} N(R,R'),
\]
where \( N(R,R') \) is the number of topplings at site R due to an avalanche triggered by adding a grain at site R' in configuration \( C \).

This function highlights which regions of the lattice are most frequently involved in avalanche propagation.

\begin{figure}[htbp]
    \centering
    \begin{minipage}[t]{0.48\textwidth}
        \centering
        \includegraphics[height=0.55\textwidth]{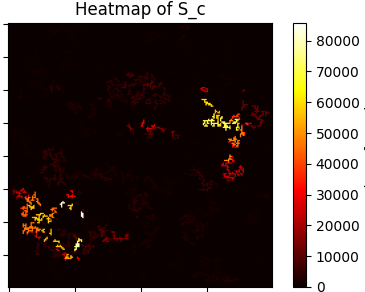}
        \caption{The heatmap of function $s$ for a configuration of lattice size 200}
        \label{fig:f1_fit}
    \end{minipage}%
    \hfill
    \begin{minipage}[t]{0.48\textwidth}
        \centering
        \includegraphics[height=0.55\textwidth]{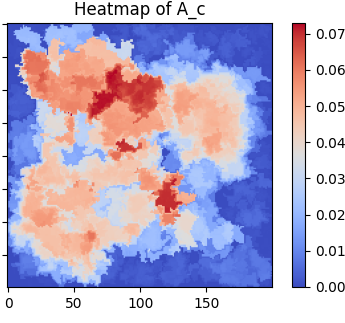}
        \caption{The heatmap of function $A$ for the same configuration, normalised by $L^2$}
        \label{fig:f2_fit}
    \end{minipage}
\end{figure}

Previous analyses have found that the expectation of \(s\) scales as approximately \(0.08\,L^2\). We confirm this behavior in our simulations by computing the lattice average of \(A\), and we find

\[
\langle A \rangle \approx 0.085\, L^2.
\]

\begin{figure}[htbp]
    \centering
    \includegraphics[width=0.7\textwidth]{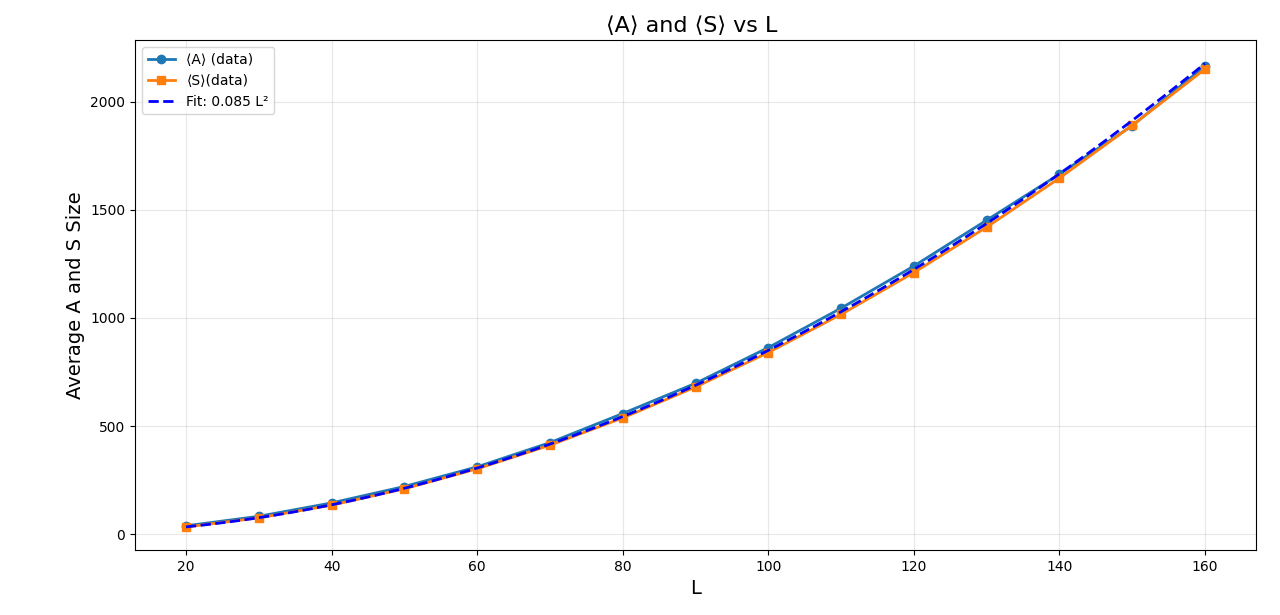}
    \caption{Plot of the expectation values $\langle A\rangle$ and $\langle s\rangle$ as a function of $L$, verifying that $\langle A\rangle=\langle s\rangle$.
}
    \label{fig:f1_fit}
\end{figure}

\section{Numerical Procedure and Details of Simulations}

Simulations were performed on lattice sizes ranging from \( L = 20 \) to \( 160 \) on open boundary conditions. For each size, the system was allowed to reach a critical state and then averaged over 10,000 configurations to compute \( A(R) \), and subsequently the empirical probabiliy distribution \( P(A = a,L) \).
To determine $A(R)$, we fix the configuration $C$, and add a particle at $R'$,counting the number of topplings at R, then reset to configuration $C$, add at a different site, and sum the activity at $R$ over all points of addition.

We analyze the probability distribution \( P(A = a,L) \), where \( A(R) \) denotes the number of topplings at site R=\( (i,j) \), accumulated over many randomly initiated avalanches. Data were collected for system sizes ranging from \( L = 20 \) to \( L = 160 \), averaging over \( 10^4 \) configurations per size.

We plot the probability distribution \( P(A = a,L) \) for small avalanche activity \( a \),and we find that for fixed values of \( a \ll L^2 \), the probability scales inversely with the linear system size:

\[
P(A = a, L) \sim \frac{g(a)}{L}.
\]

\begin{figure}[h!]
    \centering
    \includegraphics[width=0.6\textwidth]{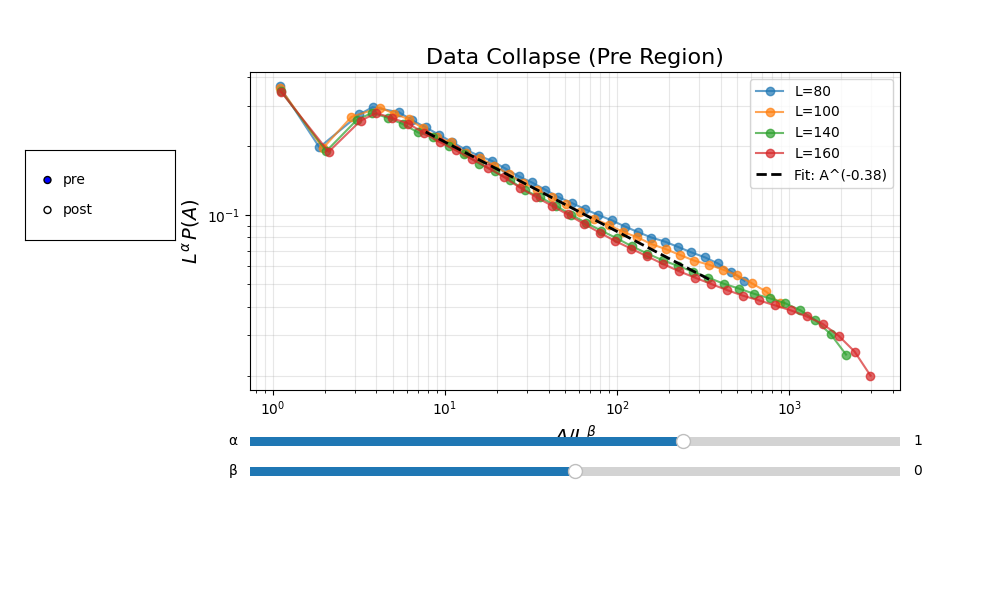} 
    \caption{Log–log plot of $g(a)=LP(A,L)$ versus $a$ for small $a$. The slope of the straight line indicates $g(a) \sim a^{-0.385\pm{0.03}}$. $R^2=0.9839$ for fit. The data can also be fitted fairly well with the form $g(x) \approx  c_1 \, x^{-1/2}+c_2$ where $x$ is $a/L^2$ and $c_1,c_2$ are constants numerically determined.}
    \label{fig:g_small_a}
\end{figure}

\begin{figure}[H]
    \centering
    \includegraphics[width=0.6\textwidth]{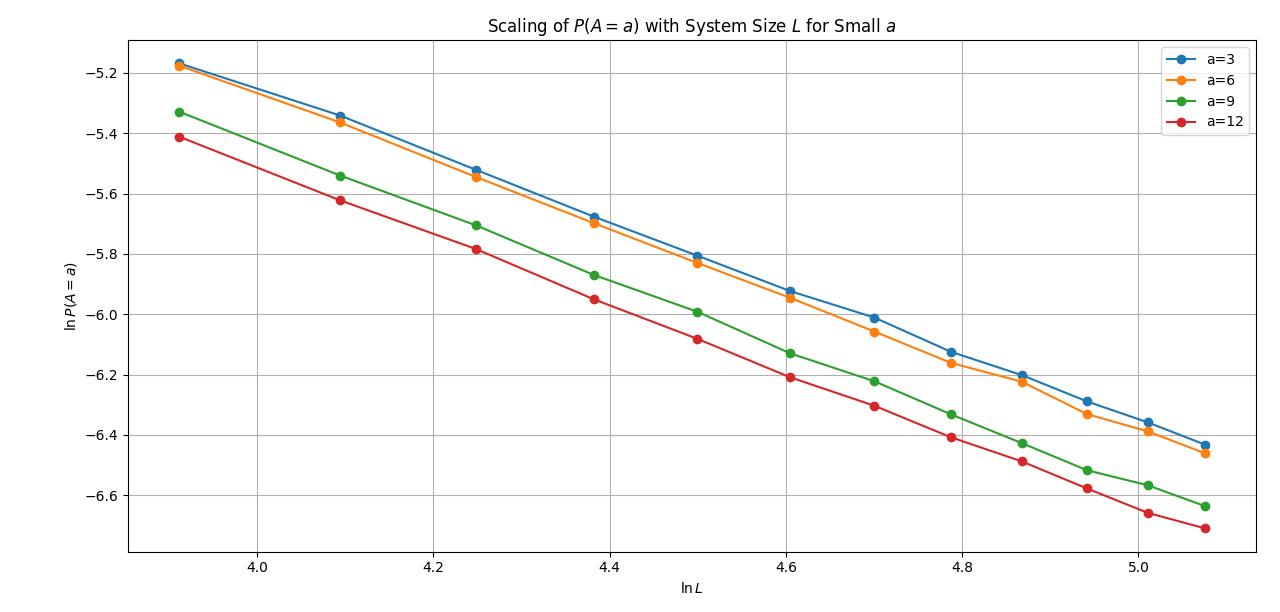}
    \caption{Log-log plot of \( P(A = a) \) vs. \( L \) for small fixed values of \( a \). The observed slope of \(-1\) indicates \( P(A = a) \sim 1/L \).}
    \label{fig:small_a_loglog}
\end{figure}

This trend is illustrated in Fig.~\ref{fig:small_a_loglog}, where we plot \( \log P(A = a) \) against \( \log L \) for small fixed values of \( a \). The slope of approximately $-1\pm0.015$ confirms the \( 1/L \) dependence.

\subsection*{Unified Scaling Form Across Avalanche Sizes}

The full distribution is well described by a single finite-size scaling form:

\[
\boxed{
P(A, L) \sim \frac{1}{L^2} F\Big(\frac{A}{L^2}\Big), \quad F(u) = \big(c_1 u^{-0.5}+c_2\big) e^{-c_3 u-c_4u^2},
}
\]

where \(c_1, c_2, c_3,c_4\) are constants determined from fits to simulation data.

\begin{figure}[H]
    \centering
    \begin{minipage}[t]{0.49\textwidth}
        \centering
        \includegraphics[width=\textwidth,height=0.65\textwidth,keepaspectratio=false]{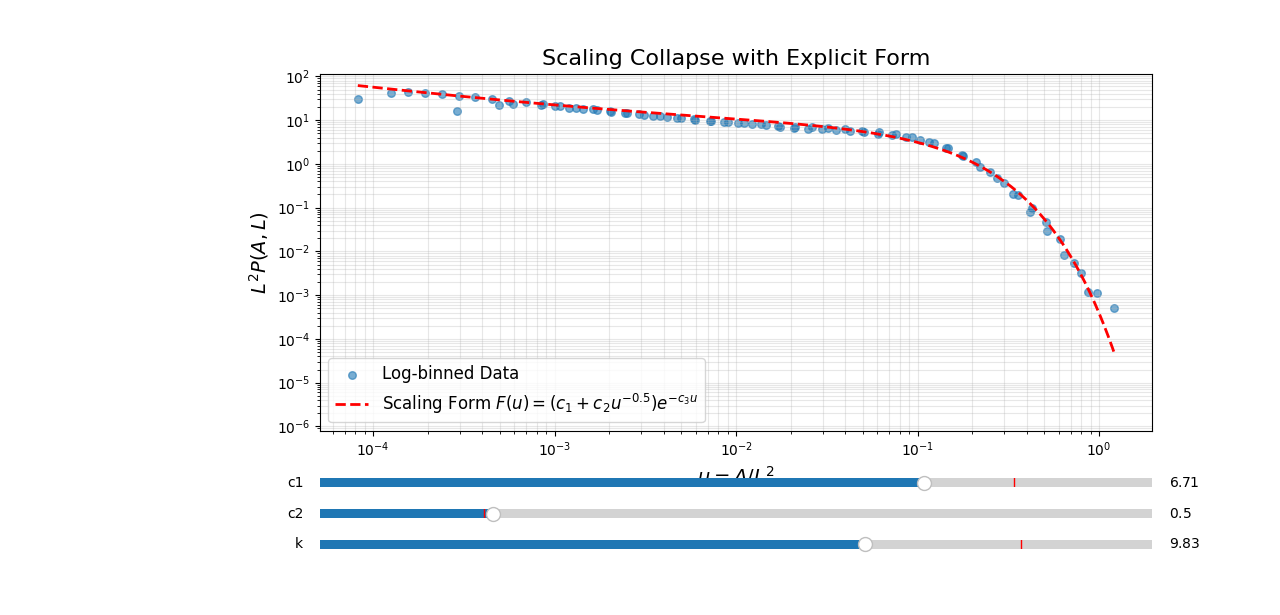}
        \caption{Scaling collapse of \(P(A,L)\) with the fitted form, log binned.}
        \label{fig:F_scaling_collapse}
    \end{minipage}\hfill
    \begin{minipage}[t]{0.49\textwidth}
        \centering
        \includegraphics[width=\textwidth,height=0.65\textwidth,keepaspectratio=false]{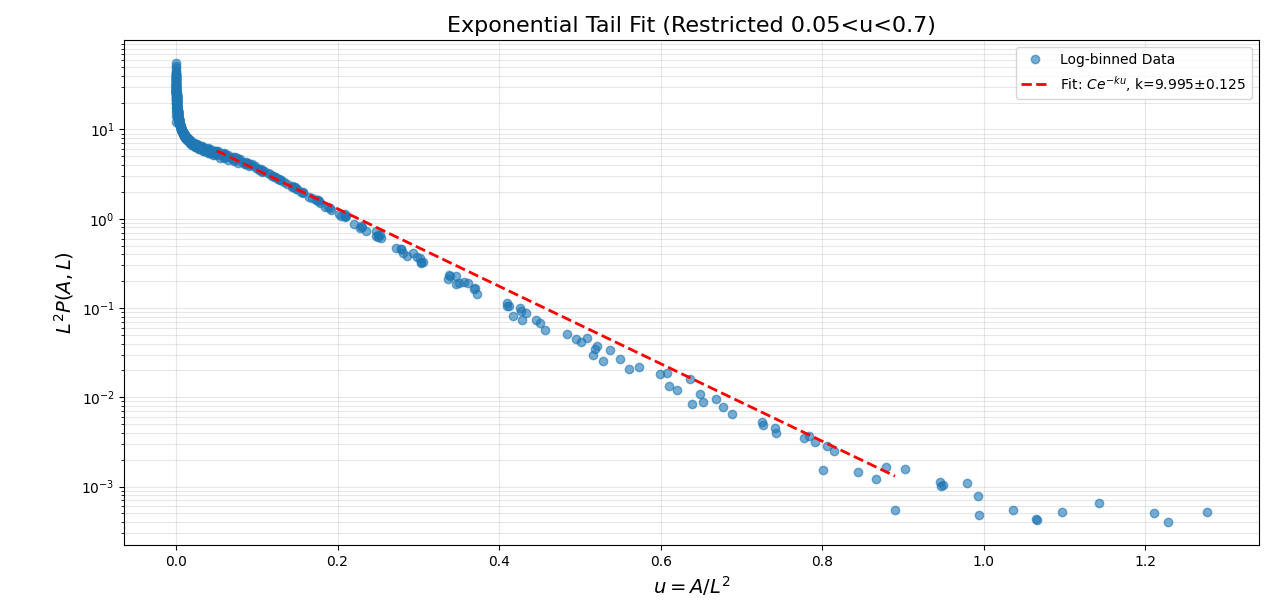}
        \caption{Semilog plot of \(P(A,L)\) vs.\ \(A\); linear for \(A>0.1L^2\).}
        \label{fig:semilog_PA}
    \end{minipage}
\end{figure}

\par
For small avalanches (\( A \ll L^2 \)), the argument \( u = A/L^2 \ll 1 \), and the scaling function is dominated by the term \( c_1 u^{-0.5}+c_2 \). Substituting \( u = A/L^2 \) gives

\[
F(u) \sim c_1 u^{-0.5}+c_2 = c_1 \frac{L}{A^{0.5}}+c_2.
\]

Thus, the probability becomes

\[
P(A, L) \sim \frac{1}{L^2} \cdot F(u)  \sim \frac{1}{L} \cdot \frac{c_1}{A^{0.5}}+\frac{c_2}{L^2},
\]

recovering the observed \( 1/L \) scaling for small avalanche sizes. The small constant \( c_2 \) in \( F(u) \) is included to account for the observed slope \( \sim -0.385 \) in the small-\( A \) regime.  

\par
For large avalanches (\( A \sim \mathcal{O}(L^2) \)), the exponential term \( e^{-c_3 u-c_4u^2} \) dominates, ensuring a rapid decay of \( P(A,L) \)

This unified scaling form therefore captures both the small-\( A \) \( 1/L \) trend and the large-\( A \) bulk suppression within a single analytic expression.

The scaling analysis of the activity distribution $P(A;L)$ reveals a crossover in its
functional form. As shown in Fig.~5, the distribution follows a power-law-like scaling for small
$A$ but undergoes a  crossover to an exponential decay for larger avalanche activities.This crossover occurs at a specific scale 
$A^* \sim 0.1 L^2$. In other words, there exists a crossover scale $A^*$ such that for 
$A \ll A^*$ the distribution exhibits power-law behavior, while for $A \gg A^*$ it is dominated 
by exponential decay.
A key finding of this work is that for a fixed A, Prob(A) goes to 0 as L tends to $\infty$
\begin{figure}[htbp]
    \centering
    \includegraphics[width=0.7\textwidth]{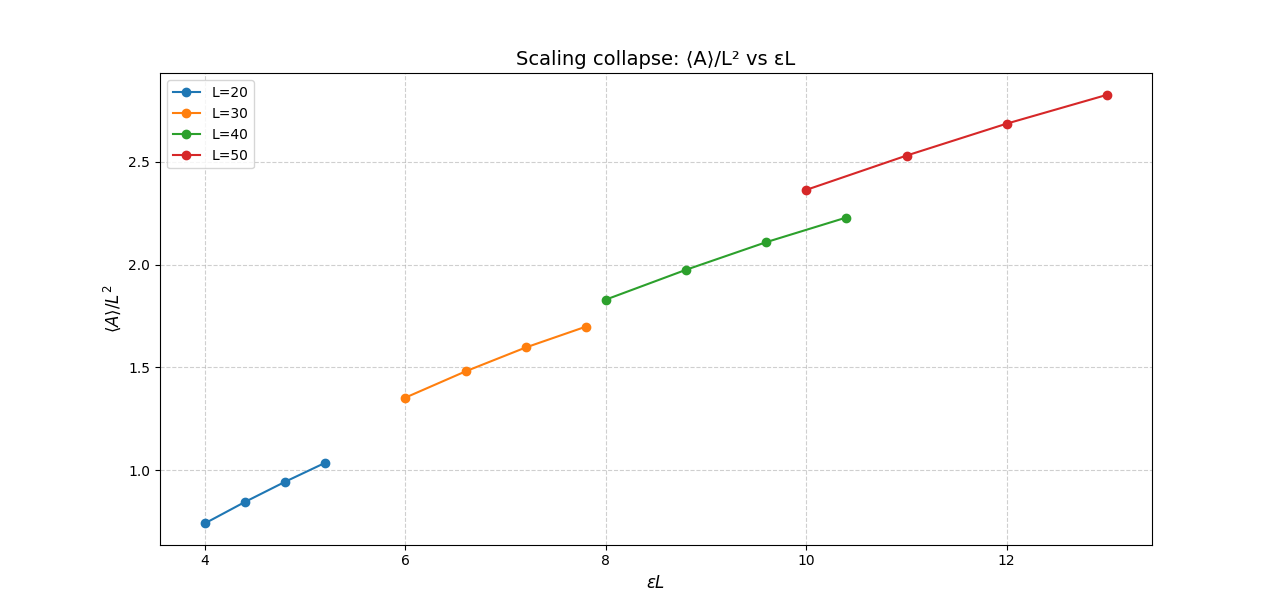}
    \caption{Plot of the $\langle A\rangle/L^2$ and  as a function of $\epsilon L$ , indicating $\langle A\rangle \sim \epsilon L^3$ as the leading order term.
}
    \label{fig:f1_fit}
\end{figure}

 Consider a rare configuration $C_\epsilon$ in an $L \times L$ Abelian Sandpile Model (ASM), where the total mass of the configuration exceeds from C by an amount $\epsilon L^2$ without changing the no of active height 3 sites.

We can see via simulations by generating such configurations $C_\epsilon$ that 
\[
\langle A \rangle_C \sim \epsilon L^3.
\]

We can see that the fractional number of states $C_\epsilon \sim \exp\big(-\epsilon^2 L^2\big)$

This gives us  $F(x \sim \epsilon L) \gtrsim \exp(-x^2), \text{ for large } x \sim \epsilon L$

\section{Scaling Theory}

We start by assuming a general scaling form for the avalanche size distribution \(P(A,L)\):
\[
P(A = a; L) \sim L^{-\alpha} F\Big(\frac{a}{L^\beta}\Big),
\]
where \(\alpha\) and \(\beta\) are scaling exponents to be determined, and \(F(u)\) is a scaling function of the dimensionless variable \(u = a / L^\beta\).

\paragraph{Normalization Condition:} 
The total probability must satisfy
\[
\sum_a P(A = a; L) = 1.
\]
Substituting the scaling form and approximating the sum by an integral for large \(L\):
\[
\sum_a P(A = a; L) \sim \int_0^\infty L^{-\alpha} F\left(\frac{a}{L^\beta}\right) \, da.
\]
Changing variables \(u = a / L^\beta \implies da = L^\beta du\):
\[
\sum_a P(A = a; L) \sim \int_0^\infty L^{-\alpha} F(u) \, L^\beta du = L^{\beta - \alpha} \int_0^\infty F(u) \, du.
\]
For normalization to hold independently of \(L\), we must have
\[
\alpha = \beta.
\]

\paragraph{Expectation Value:} 
The expected avalanche size scales as
\[
\langle A \rangle = \sum_a a \, P(A = a; L) \sim \int_0^\infty a \, L^{-\alpha} F\left(\frac{a}{L^\beta}\right) \, da.
\]
Changing variables \(u = a / L^\beta \implies da = L^\beta du\):
\[
\langle A \rangle \sim \int_0^\infty (L^\beta u) \, L^{-\alpha} F(u) \, L^\beta du = L^{2\beta - \alpha} \int_0^\infty u F(u) \, du.
\]
Since theory and simulations indicate \(\langle A \rangle \sim L^2\), we require
\[
2\beta - \alpha = 2 \quad \implies \quad \beta = 2, \, \alpha = 2.
\]

\paragraph{Small- and Large-\(u\) Behavior:} 
The scaling function \(F(u)\) controls the shape of the distribution:

\begin{itemize}
    \item For small \(u \lesssim 0.1\), assuming \(F(u) \sim u^{-1/2}\), we obtain \(P(A) \sim 1/L\).  
    \item For large values $u \gtrsim 0.1$, numerics suggest that 
$F(u) \sim \exp(-c_3 u - c_4 u^2)$, with $c_3 \sim 10$ and 
$c_4 = O(1)$. As $\epsilon$ approaches the critical value such that all 
sites have height $3$, the fraction of such configurations is of order 
$\exp(-c_4 u^2)$, indicating that $c_4 = O(1)$.

\end{itemize}

\section{Discussion}

Our results indicate that the avalanche activity distribution \(P(A = a)\) is best described by a single scaling form of the type
\[
P(A = a; L) \sim \frac{1}{L^2} F\Big(\frac{a}{L^2}\Big),
\]
where the scaling function \(F(u)\) captures both the small- and large-avalanche behavior. A suitable form that accounts for the observed piecewise behavior is
\[
F(u) = \big(c_1 u^{-0.5}+c_2\big) e^{-c_3 u-c_4u^2}.
\]
In this formulation, the first term dominates for small \(u\) (\(a \ll L^2\)), reproducing the effective \(1/L\) scaling observed in the boundary-driven small-avalanche regime, while the exponential term governs the large-\(u\) (\(a \sim L^2\)) decay, corresponding to the bulk-dominated avalanches.

For small avalanches, the data collapse reveals that
\[
L^2 P(A; L) \sim c_1 \left(\frac{a}{L^2}\right)^{-0.5}+c_2 \quad \Rightarrow \quad P(A; L) \sim \frac{1}{L} a^{-0.385},
\]
consistent with the observed slope in the pre-cutoff region. This demonstrates that boundary effects generate a scale-free contribution, yet the system-wide average is dominated by larger avalanches.

Studying the site activity $A(R)$, rather than the conventional avalanche size $s(R)$, is crucial because it shifts the focus from global avalanche statistics to local participation patterns. While $s(R)$ tells us how large an avalanche starts at a site, $A(R)$ measures how often a site participates in avalanches anywhere in the system—essentially mapping the \emph{hotspots} of avalanche propagation. This provides fundamentally different information about the internal dynamics of self-organized criticality, characterizing which regions serve as frequent conduits for activity spread. Understanding this local activity landscape is essential for connecting sandpile models to real-world systems where only partial observations are possible—such as neural recording electrodes or local strain sensors—and may ultimately help decode how global criticality emerges from local interacting thresholds.

Overall, the scaling form provides a unified description that satisfies both normalization and the expected \(\langle A \rangle \sim L^2\), while accurately capturing the observed distribution across all avalanche sizes.

\begin{figure}[H]
    \centering
    \includegraphics[width=0.7\textwidth]{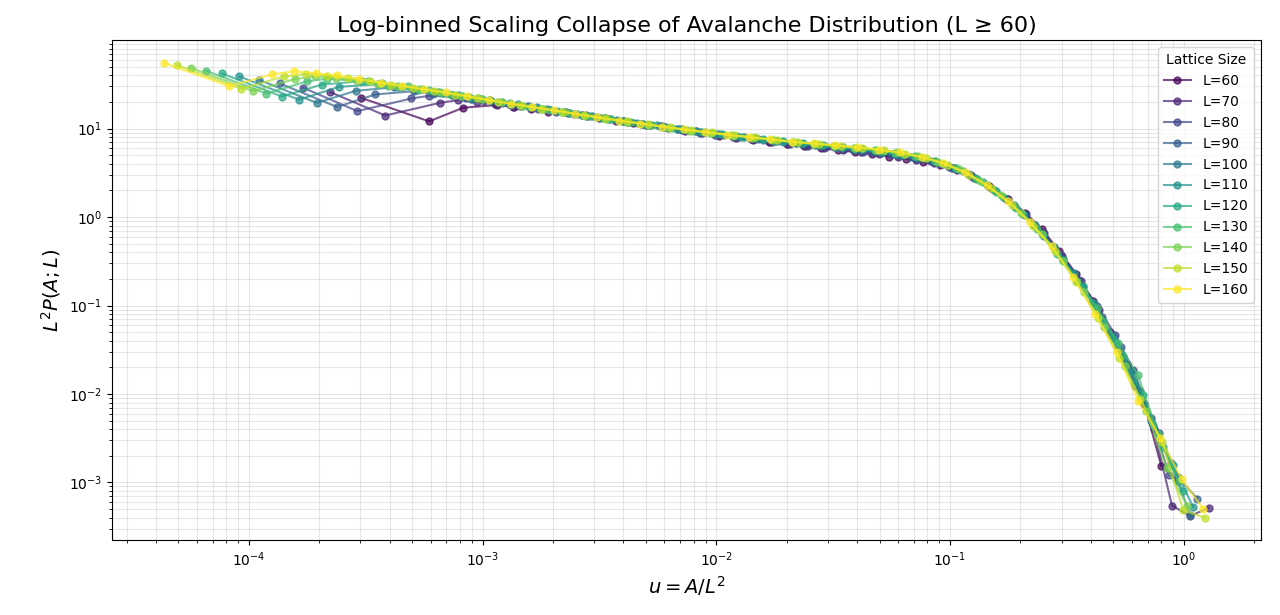}
    \caption{Scaling collapse of \( P(A,L) \) for multiple system sizes using log binned data}
    \label{fig:F_scaling_collapse}
\end{figure}
.

\noindent
We note some significant deviations from the asymptotic behavior. In particular, the probability 
of exactly $A=2$ topplings shows a pronounced dip, reflecting combinatorial constraints 
in the allowed toppling sequences.

\section*{Acknowledgments}
I would like to express my sincere gratitude to Dr.~Deepak Dhar for his insightful discussions and invaluable guidance throughout the course of this project. I also gratefully acknowledge the support and stimulating research environment provided by ICTS--TIFR, and access to the MARIO cluster.

\bibliographystyle{plain}

\end{document}